\documentclass[showpacs,preprintnumbers,amsmath,amssymb,twocolumn]{revtex4}
\usepackage{graphicx}
\usepackage{dcolumn}
\usepackage{bm}

\begin{document}
\title{Dynamic screening and plasmon spectrum in bilayer graphene}
\author{Wen-Long You}
\affiliation{School of Physical Science and Technology, Soochow
University, Suzhou, Jiangsu 215006, People's Republic of China}
\author{Xue-Feng Wang}
\email{xf_wang1969@yahoo.com}
\affiliation{School of Physical Science and Technology, Soochow
University, Suzhou, Jiangsu 215006, People's Republic of China}

\begin{abstract}
We have theoretically studied the collective response properties of
the two-dimensional chiral electron gas in bilayer graphene within
the random phase approximation. The cooperation of external
controlling factors like perpendicular electric bias, temperature,
doping, and substrate background provides great freedom to manipulate
the dynamic dielectric function and the low-energy plasmon
dispersion of the system. Intriguing situations with potential
application are systematically explored and discussed. Extra
undamped plasmon modes might emerge under electric bias. They have
almost zero group velocities and are easy to manipulate.
\end{abstract}
\maketitle

\section{INTRODUCTION}
Experimental breakthrough in isolation of high-quality few-layer
graphene by exfoliation and epitaxial growth has led to intense
experimental and theoretical interest in graphene materials
\cite{review}. Recently, much
interest has been focused on the AB-Bernal stacked bilayer graphene (BLG)
for fundamental physics and application potential in nanotechnology
\cite{oost,zhan,mcca1,min,gava,stau,sarm,sens,gama,
wang1,wang2,borg1,borg2,
stau2,nand,weit,mcca,novo,kats,wrig,prec,vela}. With its own special
nature, BLG inherits some characteristics of the monolayer graphene
(MLG) carrying chiral Dirac fermions. Intrinsic BLG is identified as
a zero-gap semiconductor with quadratic band dispersion in
low-energy regime instead of the linear band dispersion in MLG. In
such a case, the density of state (DOS) in BLG at the two
nonequivalent Dirac points, called the $K$ and ${K^\prime}$ points,
is a constant, in contract to vanishing DOS in MLG. Hence, BLG
shares some similar features with two-dimensional electron gas
(2DEG). Specifically, an energy gap between the conduction and
valence bands can be easily opened and tuned by introducing an
electrostatic potential bias between the two graphene layers
\cite{oost,zhan,mcca1,min,gava}. The bias modifies the parabolic
band structure, and increases the DOS at the top of the valance band
and the bottom of conduction band.

Previous studies have clearly evidenced that Coulomb interactions play a
significant role in graphene \cite{Elias,ZQLi}. Similarly,
electron-electron interactions in BLG can also lead to exotic
phenomena. The many-body effects are crucial to understanding the
transport and optical properties of the system. A particularly
fruitful phenomenon is the dynamic screening. The frequency
dependent screening determines the elementary quasiparticle spectra
as well as the collective modes. At zero temperature, the screening
properties and the plasmon spectrum in BLG have been studied
analytically in the two- and four-band approximations
\cite{stau,sarm,sens,gama}. For systems at finite temperature and under electric bias,
numerical calculations have been employed to study the properties in
the two-band approximation \cite{wang1,wang2}. Because of the
distinguished energy band dispersions and chiralities, Coulomb screening
properties and collective excitations in BLG exhibit significantly
different behavior from the MLG and conventional 2DEGs.
Experimentally the plasmon spectra and damping properties in
graphene structures have been studied by methods such as
electron-energy-loss spectroscopy (EELS) \cite{bost,kram,liu,lu}.

Surface plasmon modes have been used for the rapidly developing
terahertz (THz) technology \cite{wangkl,maie} and stimulated
emission of plasmon in graphene has been proposed to be used as THz
laser at room temperature \cite{rana}. One efficient way to increase
the net plasmon gain is to decrease the group velocity of plasmon in
the system. The plasmon dispersion in BLG is similar to that in MLG
and might be easier to manipulate with the help of applied electric
bias between the two layers. In this paper, based on Ref.[\onlinecite{wang2}],
we numerically study the
dynamic screening properties of the Coulomb interaction in BLG
systems within the random phase approximation (RPA)
and consider a fair general situation in which both the temperature
and the bias voltage are finite.
We calculate the dielectric function of BLG
$\epsilon(\textbf{q},\omega)$ at arbitrary wavevectors $\textbf{q}$
and frequency $\omega$. The zeros of the real part $\epsilon_r$ give
the dispersion of the plasmon modes; the imaginary part $\epsilon_i$
indicates their damping properties to single particle excitations;
the imaginary part of $1/\epsilon$ is related to their optical
spectral weight.
Beyond from Ref.[\onlinecite{wang2}], here we have explored systematically and
in a high accuracy the screening properties and the plasmon spectra in a wide
parameter space, and in some parameter regimes have found
two extra plasmon modes with almost zero group velocities.

\section{MODEL}
Since there are four inequivalent carbon atoms, the BLG system should
be described by the four-band model which gives a hyperbolic
dispersion \cite{mcca}. In the four-band model, a split-off
band is located at $\Delta \approx 0.4$ eV above the lowest
conduction band and a sombrero shape of
band is formed at the bottom of the conduction band in the existence of electric bias.
If the Fermi energy
($\approx 45.4$ meV under $U=60$ meV for $n=10^{12}$ cm$^{-2}$) plus
the thermal energy ($\approx 26$ meV at $T=300$ K) is much lower
than $\Delta$, the effect from the split-off band should be
negligible. For a bias potential $U=60$ meV, the height of the
sombrero is $\delta U=0.5U(1-\Delta/\sqrt{U^2+\Delta^2})\approx
0.3$ meV \cite{mcca1,mcca}.
In systems with carrier density $n$ higher than the critical density
$n_c = U^2/(\pi \hbar^2 v_F^2)$
($0.27 \times 10^{12}$ cm$^{-2}$ for $U$ = 60 meV), the Fermi energy is above
the sombrero and the topology of the Fermi surface is not affected by the sombrero.
In this case, the effect of the sombrero appears as a DOS broadening
(with a width of the height of the sombrero) at the bottom of the conduction band.
Since one part of the thermal effect is also similar to a DOS broadening with a width
of the thermal energy, the effect of the sombrero can also be neglected if the thermal
energy is much higher than the sombrero height.
Therefore, the low-energy properties of the system can be
qualitatively well characterized by the two-band parabolic
approximation which is valid in the range of density $n$ (0.5 to 2
$\times 10^{12}$ cm$^{-2}$), temperature $T$ (4.2 to 300 K), and
bias potential $U$ (up to 60 meV) most interested in this study.
Within such a picture, a pair of chiral
parabolic electron and hole bands touch each other at the Dirac (or
the charge neutrality) point, and each band has a four-fold degeneracy
arising from spin and valley degrees of freedom.
Note that the quadratic band dispersion in the two-band approximation
deviates from the hyperbolic band dispersion predicted by the four-band
model even in the intermediate energy regime. The corresponding results
in some situations might differ quantitatively from those in real systems.
For the sake of completeness, some zero-temperature results of systems
with low carrier density, where the sombrero effect might not be negligible,
are presented in the following.

The low-energy effective Hamiltonian describing electrons of
moderate energies in the $K$ valley of biased BLG reads as \cite{mcca}
\begin{eqnarray}
H_K = \frac{\hbar^2}{2m^*} \left(\begin{array}{c c} 0& k_{-}^2 \\
k_{+}^2 & 0
\end{array} \right) + \frac{U}{2} \left(\begin{array}{c c} 1 & 0 \\
0 & -1
\end{array} \right)
\end{eqnarray}
In the first term, the wavevector $\textbf{k}=({k_x,k_y})$ with
$k_{\pm} = k_x \pm i k_y$ is measured from the $K$ point; the
effective mass is $ m^* = \gamma/(2v_F^2)\approx 0.035m_0$ with
$\gamma$ the interlayer tunneling amplitude inherent in the BLG
system, $v_F$ the graphene Fermi velocity, and $m_0$ the free
electron mass. The second term arises from the electrostatic
potential bias $U$ between the two graphene layers separated by a
distance $d=3.35 {\AA}$. The eigenenergy and eigenwavefunction of
the above Hamiltonian read \cite{wang2}
\begin{eqnarray}
E_\textbf{k}^\lambda &=& \lambda \sqrt{(\frac{\hbar^2k^2}{2m^*})^2 +
(\frac{U}{2})^2}, \label{energy}\\
\Psi_\textbf{k}^{\lambda } &=& e^{i\textbf{k}\cdot\textbf{r}} \left(
\begin{array}{c} \sin ( \frac{\alpha_\textbf{k}}{2} + \frac{  1+\lambda
}{4}\pi)
\\ -\cos(\frac{\alpha_\textbf{k}}{2} + \frac{1+\lambda }{4}\pi)
e^{i2\theta_\textbf{k}}
\end{array}\right),
\end{eqnarray}
where $\lambda = \pm 1$ denotes respectively the conduction and the
valence band. Here $\theta_\textbf{k}$ is the azimuth of the vector
$\textbf{k}$, i.e., $\tan \theta_\textbf{k}$ =$k_y /k_x$, $k=|\textbf{k}|$, and
$\alpha_\textbf{k}$ indicates the ratio of the kinetic energy to the
potential bias with $\tan \alpha_\textbf{k}$ = $\hbar^2 k^2/( m^{*}
U)$. For such an energy dispersion the DOS of the system is
\begin{eqnarray}
D(E)=\frac{g_\nu}{(2\pi)^2} \int \frac{dS}{\vert \nabla_\textbf{k}
E\vert} = \frac{m^*}{\pi \hbar^2} \frac{E}{\sqrt{E^2
-(\frac{U}{2})^2}},
\end{eqnarray}
where $g_\nu$ is a constant degeneracy factor. Here $g_\nu=4$ comes
from the degenerate two spins and two valleys at $K$ and
${K^\prime}$. Under finite bias $U$, the DOS of the BLG diverges on
the edge of the energy gap $E=|U/2|$. It has been shown that the
static dielectric constant at $\textbf{q}=0$ in intrinsic BLG is
much larger than that in MLG due to the presence of a finite
DOS
at the $K$ point \cite{wang1,wang2}. The bias might further modify
the system's properties in a large range of variety.

The wavevector and frequency-dependent dielectric function $\epsilon
(\textbf{q}, \omega)$ tells the response of the system to a weak
external perturbation, and determines a variety of many-body related
terms such as self-energy, carrier lifetime, and mobility, as well
as characterization of other excitations. In the RPA, it is given as
\begin{eqnarray}
\epsilon (\textbf{q}, \omega) = 1 -  v_q \Pi(\textbf{q}, \omega)
\label{epsilon},
\end{eqnarray}
where $v_q = e^2/(2\varepsilon_0 \varepsilon_b q)$
(with the background dielectric constant $\varepsilon_b$) is the Fourier
transformation of the bare Coulomb interaction and the electron-hole
propagator is originated from the bare bubble diagram \cite{wang2}
\begin{eqnarray}
\Pi(\textbf{q},\omega) = 4\sum_{\lambda,\lambda^{\prime},\textbf{k}}
\vert g_\textbf{k}^{\lambda,\lambda^{\prime}}(\textbf{q})\vert^2
\frac{f(E_{\textbf{k}+\textbf{q}}^{\lambda ^\prime}) -
f(E_\textbf{k}^\lambda)}{\omega +
E_{\textbf{k}+\textbf{q}}^{\lambda^{\prime}} -
E_{\textbf{k}}^\lambda + i \delta }.
\end{eqnarray}
Here $f(x)$ is the Fermi function and the vertex factor reads
\begin{eqnarray}
\vert g_\textbf{k}^{\lambda,\lambda^{\prime}}(\textbf{q})\vert^2 &=&
\vert \langle
\textbf{k}+\textbf{q},\lambda^{\prime} \vert e^{i \textbf{q}\cdot \textbf{r}}
\vert \textbf{k}, \lambda \rangle \vert^2 \nonumber \\
&=& \frac{1}{2} \left[1+\lambda \lambda^{\prime} \cos
\alpha_\textbf{k} \cos \alpha_{\textbf{k}+\textbf{q}} \right.
\nonumber \\ &+&\left. \lambda \lambda^{\prime} \sin
\alpha_\textbf{k} \sin \alpha_{\textbf{k}+\textbf{q}} \cos (2
\theta_\textbf{k} - 2\theta_{\textbf{k}+\textbf{q}}) \right].
\end{eqnarray}
When $\textbf{q}=0$ and $\textbf{q}=-2\textbf{k}$, $\vert
g_\textbf{k}^{\lambda,\lambda^{\prime}}(\textbf{q})\vert^2$ =
$\frac{1}{2} (1+\lambda \lambda^{\prime})$. Similar to unbiased BLG,
the interband vertical and back scatterings are both forbidden but
the intraband back scattering is allowed in biased BLG.
For intraband scattering with $\textbf{k}\perp \textbf{k+q}$, we have
$\vert g_\textbf{k}^{\lambda,\lambda^{\prime}}(\textbf{q})\vert^2
=\frac{1}{2} [1+\cos(\alpha_\textbf{k}+\alpha_{\textbf{k}+\textbf{q}})]$,
which becomes zero in unbiased BLG \cite{wang2}.

\section{result and discussions}

\begin{figure}
\includegraphics[width=8.5cm]{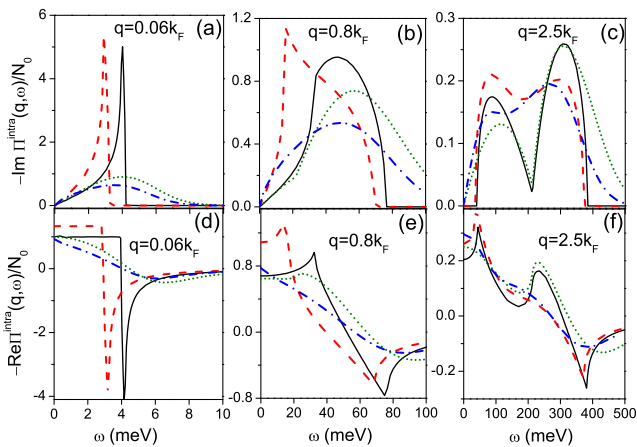}
\caption{The negative intraband electron-hole propagator in unit
of $N_0$ versus frequency $\omega$ in unbiased/biased BLG at zero/room
temperature. The imaginary part for $q = 0.06$, $0.8$, and $2.5 k_F$
is plotted in panel (a), (b), and (c), respectively, and the corresponding
real part in (d), (e), and (f).
The electron density is $n=10^{12}$ cm$^{-2}$ with the Fermi wavevector
$k_F=1.77 \times 10^8$ m$^{-1}$.
The corresponding Fermi energy is $E_F=34$ meV at $U=0$ and $E_F=45.4$ meV
at $U=60$ meV. Solid curves are for $T=0$, $U=0$; dashed $T=0$, $U=60$ meV;
dotted $T=300$ K, $U=0$; dash-dotted $T=300$ K, $U=60$ meV.}
\label{fig:intra-U}
\end{figure}

The electron-hole propagator is composed of intra-
($\lambda=\lambda'$) and inter-band ($\lambda=-\lambda'$)
components. The intraband component is expected to be similar to
that in conventional 2DEGs except the effect of chirality and
deformation of energy band.
The interband one,
which can be manipulated by the bias voltage,
modifies qualitatively the screening properties of the system.
Since the approximate energy spectrum we use in this study is isotropic,
the obtained properties are also isotropic and we use
$q \equiv \vert \textbf{q} \vert $ and $k \equiv \vert \textbf{k} \vert$
in the following discussion.

In Fig. \ref{fig:intra-U}, we plot the negative intraband
propagator in unit of $N_0=2m^\ast/\pi$ (the DOS of intrinsic BLG)
versus frequency $\omega$ for three typical wavevector $q$ values.
Similar to the 2DEG result at zero temperature, the imaginary part
is nonzero in the single particle continuum $0<\omega<\omega_u$ for
$0<q<2k_F$ and $\omega_l<\omega<\omega_u$ for $q>2k_F$, with
$\omega_l=|E^1_{k_F-q}-E^1_{k_F}|$ and
$\omega_u=E^1_{k_F+q}-E^1_{k_F}$. The derivative of the imaginary
part is not continuous at the continuum edges and for $0<q<2k_F$ is
also not continuous at $\omega=\omega_l$ besides $\omega=0$ and
$\omega_u$. When the bias $U$ increases, the structures of the
curves shift to lower energy because the energy band is narrowed and
the group velocity of electrons at the bottom of the conduction band
decreases. When the temperature increases, the sharp edges become
smoother and the nonzero range gets wider as expected.

Different from the 2DEG result, as illustrated in
Fig.\ref{fig:intra-U}(c) for $q>k_F$ at $U=0$ and $T=0$, the
imaginary part has a sharp dip with a derivative discontinuity at
$\omega_m=q^2/2m^\ast$ between the edges
due to the chiral nature of
the wave functions \cite{sens}.
The dip and discontinuity persist at finite
temperature but become softened and disappear under finite bias
voltage.

The real part, presented in the lower panels of
Fig.\ref{fig:intra-U}, shows a sharp peak at $\omega_l$ and a sharp
dip at $\omega_u$ with an extra peak near $\omega_m$ for $q>k_F$ and
moves in a similar way as the imaginary part with bias and
temperature. However, the structure near $\omega_m$ for both the
imaginary and real part develops in a different way from those at
$\omega_l$ and $\omega_u$ when $U$ and $T$ increase. It remains at
high temperature and removes off with the bias voltage while those
at $\omega_l$ and $\omega_u$ decay with temperature. On average, the
variation range of the propagator decreases with $q$.

\begin{figure}
\includegraphics[width=8.5cm]{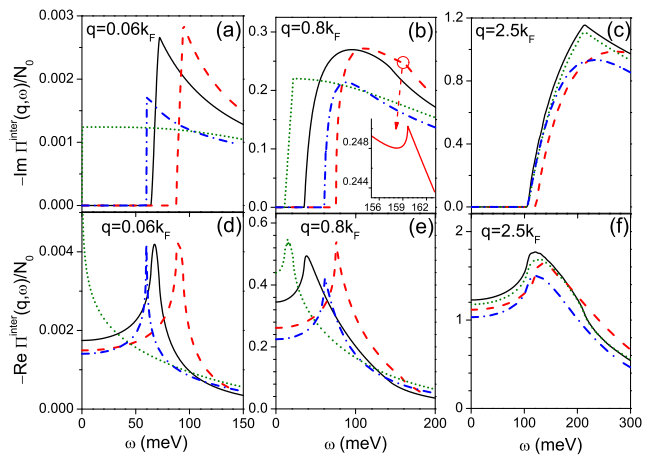}
\caption{The negative interband electron-hole propagator in unit of
$N_0$ versus frequency $\omega$ is plotted for the same parameters
set and arrangement as in Fig.\ref{fig:intra-U}.}
\label{fig:inter-U}
\end{figure}

In Fig.\ref{fig:inter-U}, we present the negative interband propagator
in unit of $N_0$ versus $\omega$ for the same three $q$ values as in
Fig.\ref{fig:intra-U}. Its imaginary part and the single particle
continuum have a minimal energy limit
$\omega_1=E^1_{k_F}-E^{-1}_{k_F-q}$ at zero temperature or
$\omega_2=2E^1_{q/2}$ at finite temperature. For small $q$
[Fig.\ref{fig:inter-U}(a)] the imaginary part increases sharply and
reaches a peak before decreases in a way $\sim 1/\omega$. For
mediate $q=0.8k_F$ [Fig.\ref{fig:inter-U}(b)] the main peak becomes
round and smooth while a sharp peak near
$\omega_3=E^1_{k_F}-E^{-1}_{k_F+q}$ appears under bias
as shown in the inset of Fig.\ref{fig:inter-U}(b).
This latter peak grows and becomes more visible as $U$ increases.
For $q>k_F$ a peak with discontinuity appears near $\omega_m=q^2/2m^*$
at zero temperature in both biased and unbiased BLG \cite{sens} as
shown in Fig.\ref{fig:inter-U}(c).

Corresponding to the continuum edges of the imaginary part, on the
curve of the real part as shown in the lower panels of
Fig.\ref{fig:inter-U}, there is a peak near $\omega_1$ ($\omega_2$)
at low (high) temperature when the electron system is degenerate
(nondegenerate). The peak near $\omega_2$ may become very sharp for
small $q$ in biased BLG because the bottom (top) of the conduction
(valence) band becomes flat and the DOS diverges on the edge of
energy gap. The overall contribution of interband excitation to the
propagator increases with $q$ and in a way $\sim q^2$ at small $q$.

\begin{figure}
\includegraphics[width=8.5cm]{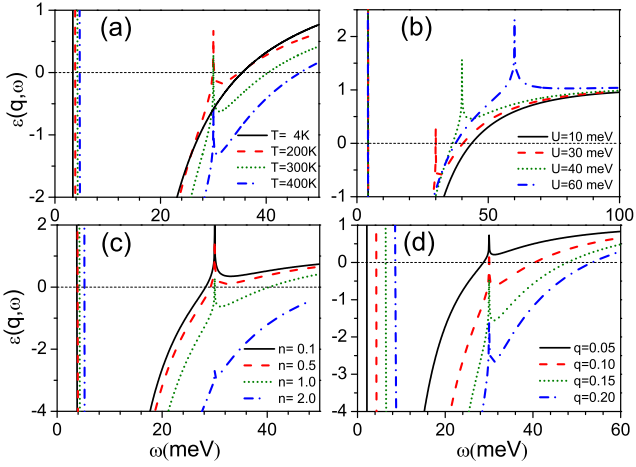}
\caption{The real part of the dielectric function $\epsilon_r$ vs
frequency $\omega$ when $\varepsilon_b=1$ in variety of bias $U$,
temperature $T$, electron density $n$ ($\times 10^{12}$ m$^{-2}$),
and wavevector $q$ ($\times 10^8$ m$^{-1}$) is plotted in (a)-(d),
respectively. Parameters $U=30$ meV, $n=10^{12}$ m$^{-2}$,
$q=0.1\times 10^8$ m$^{-1}$, $T=300$ K are used if not specified in the panels.}
\label{fig:uepr-u-t-q}
\end{figure}

As we know, the electron-hole propagator reflects the electric
polarizability of a many-body system screening a Coulomb potential.
After being reduced by the background dielectric constant
$\varepsilon_b$, it determines the dielectric function $\epsilon (q,
\omega)=\epsilon_r+i\epsilon_i$ as indicated in Eq.(\ref{epsilon}).
The zero of the real part $\epsilon_r$ gives the collective
excitation of the system in the absence of external electromagnetic
field. The imaginary part $\epsilon_i$ gives the spectrum of the
single particle excitation. In the presence of only intraband single
particle excitation as in conventional 2DEG, there exist maximally
two plasmon modes, one acoustic mode of frequency $\omega_A$ within
the single particle excitation and another optical mode $\omega_O$,
because $-$Re[$\Pi(q, \omega)$]  has only one dip below zero as
shown in Fig.\ref{fig:intra-U}. The acoustic mode is thus always
overdamped with little spectral weight and not experimentally
relevant. The contribution from interband excitation introduces fine
structures to $\epsilon_r$ near zero and at least two extra modes,
$\omega^p_3$ and $\omega^p_4$ may emerge.

In Fig.\ref{fig:uepr-u-t-q}, we display $\epsilon_r$ versus $\omega$
for various temperature $T$ (a), bias voltage $U$ (b), electron
density $n$ (c), and wavevector $q$ (d). In the high frequency
limit, the effect of polarization vanishes and $\epsilon_\infty=1$.
For intrinsic BLG where $n=0$, the intraband polarization is
negligible and $\epsilon_r>0$ at $T=0$. There is no collective mode.
In other cases, the $\epsilon_r$ versus $\omega$ curve has a deep
dip at $\omega_u$ and a peak at $\omega_1$ or/and $\omega_2$. The
competition in Eq.(\ref{epsilon}) among value one,
the intra-, and inter-band contributions to the polarizability determine its fine
features. For typical parameters, as illustrated in
Fig.\ref{fig:intra-U} and \ref{fig:inter-U}, the effect of the
intra- (inter-) band polarization decreases (increases) with $q$ so
we expect that $\epsilon_r$ mainly shows intra- (inter-) band
characteristics for long (short) wavelength. In this study we are
mostly interested in the screening and collective excitation
properties of long wavelength in the system, and in
Fig.\ref{fig:uepr-u-t-q} we present the details only for $q\ll k_F$
near $\epsilon_r=0$.

The result for a system of $U=30$
meV, $n=10^{12}$ cm$^{-2}$, and $q=0.1\times 10^8$ m$^{-1}$ is
exhibited in Fig.\ref{fig:uepr-u-t-q}(a).
At low temperature $T=4$ K (solid) when the system is
degenerate, the intraband contribution dominates and a peak appear
near $\omega_1=E^1_{k_F}-E^{-1}_{k_F-q} \simeq 2E^1_{k_F}=74.4$ meV
which is out of the panel. When the temperature increases to 200 K
(dashed) and the system becomes nondegenerate, a sharp peak emerge
near $\omega_2=2E^1_{q/2}\simeq 30$ meV. Since this interband peak
is located just below the optical plasmon energy $\omega_O$, it can
introduce two extra roots $\omega^p_3$ and $\omega^p_4$ to the
equation $\epsilon_r=0$, i.e., two extra plasmon modes in the
system. The extra plasmon mode of lower energy $\omega^p_3$ is out
of the interband single particle continuum and can be only weakly
damped. When the temperature increases further (dotted and
dash-dotted) the enhanced intraband contribution shifts the peak
below zero and the extra plasmon modes disappear. In the same time,
$\omega_O$ increases with the temperature as more electrons (holes)
exist in the conduction (valence) band.

As shown in Fig.\ref{fig:uepr-u-t-q}(b), the bias voltage can
sensitively shift the interband peak and control the emergence of
the extra plasmon modes. These modes have energies (frequencies)
proportional to the bias voltage and group velocities close to zero.
The increase of the electron density enhances the degeneracy of the
system and reduces the amplitude of the interband peak at
$\omega_2=2E^1_{q/2}$ as plotted in Fig.\ref{fig:uepr-u-t-q}(c). The
enhanced intraband contribution at higher density also increases the
frequency of the optical plasmon mode $\omega_O$.

In Fig.\ref{fig:uepr-u-t-q}(d) we illustrate how the
$\epsilon_r$-$\omega$ curve develops with $q$ at room temperature
$T=300$ K. Overall the intra- (inter-) band contribution to
$\epsilon$ simply decreases (increases) with $q$ as shown previously
in Figs.\ref{fig:intra-U} and \ref{fig:inter-U}, but their effect on
the plasmon spectrum is more complicated due to their competition
with each other. When $q$ increases, the intraband introduced
$\epsilon_r$ dip becomes wide which usually results in the increase
of the plasmon frequency. The interband peak of $\epsilon_r$ is
located at the fixed energy $\omega_2 \simeq 30$ meV and its
amplitude increases with $q$. As a result, only when $\omega_2$ is
close to $\omega_O$, the interband contribution can affect
significantly the plasmon spectrum.

\begin{figure}
\includegraphics[width=9cm]{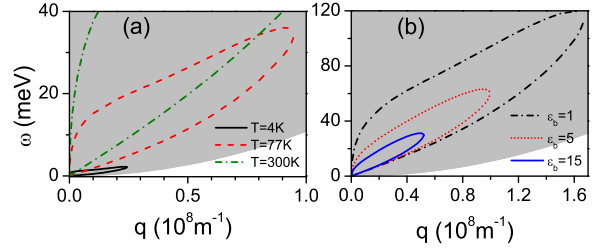}
\vspace{-3cm}
\caption{Plasmon spectrum of intrinsic BLG (a) in vacuum with
$\varepsilon_b=1$ at various temperature and (b) with various
dielectric constant at room temperature $T=300$ K. The shadow
shows the electron-hole single-particle continuum at zero temperature.}
\label{fig:u0n0-T}
\end{figure}

The plasmon spectrum in intrinsic
BLG (unbiased and undoped) depicted in Fig.\ref{fig:u0n0-T}
at various temperatures $T$ (a) and for
various background dielectric constant $\varepsilon_b$ (b). At zero
temperature, there is no carrier and no plasmon mode in the system.
At finite temperature, electrons (holes) are excited in the
conduction (valence) band and two plasmon modes emerge. In the
long-wavelength limit, their frequencies are proportional to
$\sqrt{T}$ at high temperature. Similar to the 2DEG result, we also
observe a dispersion $\omega_O \propto \sqrt{q}$  and $\omega_A
\propto q$. The background dielectric constant can also be employed
to modify the plasmon frequency as shown in Fig.\ref{fig:u0n0-T}(b).
The acoustic mode is not sensitive with $\varepsilon_b$ but the
frequency of optical mode decreases quickly with $\varepsilon_b$.

\begin{figure}
\includegraphics[width=9cm]{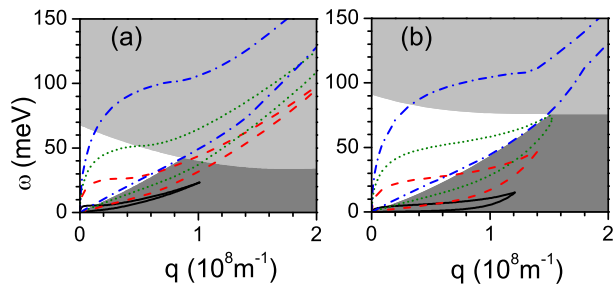}
\vspace{-8.5cm}
\caption{Plasmon spectrum at zero temperature of (a) unbiased (b)
U=60meV biased BLG in vacuum for density $n=0.1$ (solid), 0.5
(dashed), 1.0 (dotted), and 2.0 (dash-dotted) $\times 10^{12}$
cm$^{-2}$. The single-particle continuum at $n=10^{12}$ cm$^{-2}$
is also present (light/dark shadow for the inter-/intra-band part).}
\label{fig:u0-60t0-n}
\end{figure}

In Fig. \ref{fig:u0-60t0-n}, the plasmon spectrum for various
electron densities at zero temperature is illustrated in unbiased
BLG (a) and in biased one with $U=60$ meV (b). In the long-wave
limit, $q\sim 0$, the dielectric function is
dominated
by the intraband contribution. The properties of the
system at zero temperature is mainly determined by the group
velocity of electrons at the Fermi energy. The plasmon spectrum of
unbiased system is similar to that of 2DEG. If the electron density
is not high, e.g. $n\lesssim 10^{12}$ cm$^{-2}$ for $U=60$ meV as
shown in Fig.\ref{fig:u0-60t0-n}(b), the Fermi group velocity of
electrons decreases when the external electric field is turned on,
due to the band deformation, and the plasmon mode becomes softened.
With increasing $q$ the interband contribution becomes more
important, which reduces the group velocity of the optical plasmon
mode. In some cases, the group velocity of plasmon can be close to
zero, a favorite situation for the stimulated plasmon emission
\cite{rana}. For large $q$, when the optical plasmon branch enters
the interband single particle continuum, i.e. $\omega_O > \omega_1$,
the effect of interband contribution decreases and the plasmon
spectrum has a long tail in unbiased system or when the carrier
density is high. However, in biased systems with low carrier
density, the effect of the interband contribution can be significant
due to the flat band and the plasmon spectrum ends near where the
intra- and inter-band single particle continua meet at
$\omega_u=\omega_1$.

\begin{figure}[h]
\includegraphics[width=8.5cm]{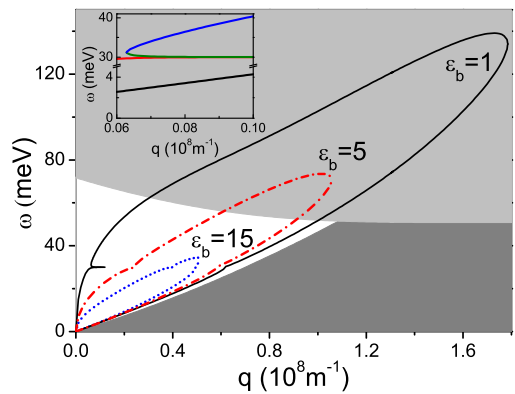}
\caption{Plasmon spectrum of biased and doped BLG with $U=30$ meV and
$n=10^{12}$ cm$^{-2}$ at room temperature $T=300$ K in environment of
different background dielectric constants $\varepsilon_b$. The inset
shows the zoomed spectrum of the plasmon branches $\omega_O$,$\omega^p_3$,
$\omega^p_4$, and $\omega_A$ for $\varepsilon_b=1$ at small $q$. The light
(dark) shadow shows the interband (intraband) single-particle continuum at
zero temperature.}
\label{fig:T300U30n1e}
\end{figure}

The competition between the intra- and inter-band contributions may
result in two extra plasmon modes for proper $U$ at finite
temperature when thermal excitation becomes important as previously
discussed in Fig.\ref{fig:uepr-u-t-q}. This is an interesting
phenomenon because
those modes have unique properties and might be used in
nanotechnology. In Fig.\ref{fig:T300U30n1e}, we plot the plasmon
spectra of a biased system with $U=30$ meV at $T=300$ K and
$n=10^{12}$ cm$^{-2}$ for background dielectric constant
$\varepsilon_b=1$ (solid), 5 (dash-dotted), and 15 (dotted). At room
temperature $T=300$ K, the electronic system is not degenerate and
the plasmon spectrum is in general similar to the one of intrinsic
BLG as shown in Fig.\ref{fig:u0n0-T}(b). However, as shown in
Fig.\ref{fig:inter-U} and \ref{fig:uepr-u-t-q}, the interband
contribution adds a sharp peak at $\omega_2$ which is around $U$ for
small $q$. As a result, the plasmon spectra are deformed at
$\omega_2$ and bifurcate in some cases where two extra plasmon modes
$\omega^p_3$ and $\omega^p_4$ emerge as shown in the inset of
Fig.\ref{fig:T300U30n1e}.

\begin{figure}
\includegraphics[width=7cm]{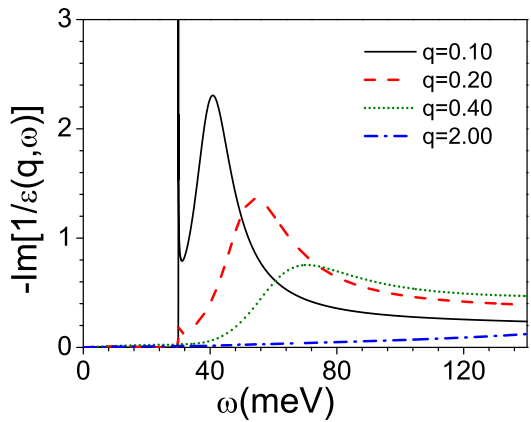}
\caption{The spectral weight function versus frequency at various
wavevectors $q$ ($\times 10^8$ m$^{-1}$) for the system with
$\varepsilon_b=1$ in Fig.\ref{fig:T300U30n1e}.}
\label{fig:invu30t300n1}
\end{figure}

The emerged two plasmon modes have almost zero group velocities and
their frequencies are proportional to the bias voltage. In addition,
their energies are in the gap of the zero-temperature single
particle continuum and the lower one is below the lower limit of
single particle excitation, $\omega_2$, at high temperature. This
suggests that the modes are undamped or weakly damped and have long
lifetime. In Fig.\ref{fig:invu30t300n1} we plot the negative
imaginary part of the dielectric function, which indicates the
spectral weight of the collective modes, for different $q$. We see
that the spectral weigh shows a negligible value (wide peak) near
the energy of the acoustic (optical) mode $\omega_A$ ($\omega_O$),
indicating that the $\omega_A$ mode is damped and the $\omega_O$ is
weakly damped. In the same time, there is a sharp and high peak at
$\omega \backsimeq U$ for small $q$ indicating that the $\omega_3$
and $\omega_4$ modes are almost undamped. These undamped plasmon
modes are similar to those in MLG but have lower group velocity.
Their energies can be easily manipulated by the bias voltage. These
undamped modes with almost zero group
velocities
then might be used in THz technology \cite{rana}.

In conclusion, we have studied systematically the many-body response
of electrons to external Coulomb perturbation and the plasmon
spectra in a biased BLG. The vertical voltage bias opens a gap
between the conduction and valence bands and increases the DOSs at
the band edges. As a result, the bias modifies the dielectric
function greatly. In the long-wave limit, a sharp and controllable
dielectric peak might appear at the energy equal to the band gap at
high temperature when the system is nondegenerate or at the energy
of double the Fermi energy at low temperature when the system is
degenerate. In some cases, two extra undamped plasmon modes appear
at energies close to the band gap energy and have almost zero group
velocities.

Wen-Long You acknowledges the support of the Natural Science
Foundation of Jiangsu Province under Grant No. 10KJB140010 and the
National Natural Science Foundation (NSFC) of China under Grant No.
11004144. Xue-Feng Wang is supported by NSFC of China (No. 11074182 and 91121021).


\begin{references}
\bibitem{review} T. Ando, J. Phys. Soc. Jpn. {\bf 74}, 777 (2005);
A. H. Castro Neto,
F. Guinea, N. M. R. Peres, K. S. Novoselov, and
A. K. Geim, Rev. Mod. Phys. {\bf 81}, 109 (2009);
D. S. L. Abergel, V. Apalkov, J. Berashevich, K. Ziegler,
and T. Chakraborty,
Adv. Phys. {\bf 59}, 261 (2010);
S. Das Sarma, S. Adam, E. H. Hwang,
and
E. Rossi, Rev. Mod. Phys. {\bf 83}, 407 (2011);
M.O. Goerbig, Rev. Mod. Phys. {\bf 83}, 1193 (2011).

\bibitem{oost} J. B. Oostinga, H. B. Heersche, X. Liu, A. F. Morpurgo, and
L. M. K. Vandersypen, Nature Mater. {\bf 7}, 151 (2007).

\bibitem{zhan} Y. Zhang, T. T. Tang, C. Girit, Z. Hao, M. C. Martin, A. Zettl M. F. Crommie,
Y. R. Shen, and F. Wang, Nature (London) {\bf 459}, 820 (2009).

\bibitem{mcca1} E. McCann, Phys. Rev. B {\bf 74}, 161403(R) (2006).

\bibitem{min} H. Min, B. Sahu, S. K. Banerjee, and A. H.
MacDonald, Phys. Rev. B {\bf 75}, 155115 (2007).

\bibitem{gava} P. Gava, M. Lazzeri, A. M. Saitta, and F. Mauri, Phys. Rev. B {\bf 79}, 165431 (2009).

\bibitem{stau} T. Stauber, N. M. R. Peres, F. Guinea, and A. H. Castro Neto,
Phys. Rev. B {\bf 75}, 115425 (2007).

\bibitem{sarm}E. H. Hwang and S. Das Sarma, Phys. Rev. Lett {\bf 101}, 156802 (2008).

\bibitem{sens}
R. Sensarma, E. H. Hwang and S. Das Sarma, Phys. Rev. B {\bf 82},
195428 (2010).

\bibitem{gama} O. V. Gamayun, Phys. Rev. B {\bf 84}, 085112 (2011).

\bibitem{wang1} X. F. Wang and T. Chakraborty, Phys. Rev. B {\bf 75}, 041404(R) (2007).

\bibitem{wang2} X. F. Wang and T. Chakraborty, Phys. Rev. B {\bf 81}, 081402(R) (2010).

\bibitem{borg1}
G. Borghi, M. Polini, R. Asgari, and A. H. MacDonald, Phys. Rev. B {\bf 80}, 241402(R) (2009).

\bibitem{borg2}
G. Borghi, M. Polini, R. Asgari, and A. H. MacDonald, Phys. Rev. B {\bf 82}, 155403 (2010).

\bibitem{stau2} T. Stauber and G. G\'{o}mez-Santos, Phys. Rev. B {\bf 85}, 075410 (2012).

\bibitem{nand} R. Nandkishore and L. Levitov, Phys. Rev. Lett. {\bf 104}, 156803 (2010).

\bibitem{weit} R. T. Weitz, M. T. Allen, B. E. Feldman, J. Martin, A. Yacoby, Science {\bf 330}, 812 (2010).

\bibitem{mcca} E. McCann and V. I. Fal'ko, Phys. Rev. Lett. {\bf 96}, 086805 (2006). 

\bibitem{novo} K. S. Novoselov, E. McCann, S. V. Morozov, V. I. Fal'ko,
M. I. Katsnelson, U. Zeitler, D. Jiang, F. Schedin,
and A. K. Geim, Nat. Phys. {\bf 2}, 177 (2006). 

\bibitem{kats}
M. I. Katsnelson, K. S. Novoselov, and A. K. Geim, Nat. Phys. {\bf 2}, 620 (2006). 

\bibitem{wrig} A. R. Wright, J. C. Cao, and C. Zhang, Phys. Rev. Lett. {\bf 103}, 207401 (2009).

\bibitem{prec} L. Prechtel, L. Song, D. Schuh, P. Ajayan, W. Wegscheider and A. W. Holleitner,
Nat. Comm. {\bf 3}, 646 (2012).

\bibitem{vela} J. Velasco Jr, L. Jing, W. Bao, Y. Lee, P. Kratz, V. Aji, M. Bockrath, C. N. Lau, C. Varma,
R. Stillwell, D. Smirnov, Fan Zhang, J. Jung and A. H. MacDonald,
Nat. Nanotech. {\bf 7}, 156 (2012).

\bibitem{Elias}D. C. Elias,
R. V. Gorbachev, A. S. Mayorov, S. V. Morozov, A. A. Zhukov, P.
Blake, L. A. Ponomarenko, I. V. Grigorieva, K. S. Novoselov, F.
Guinea and A. K. Geim, Nat. Phys. {\bf 7}, 701 (2011).

\bibitem{ZQLi}Z. Q. Li, E. A. Henriksen, Z. Jiang, Z. Hao, M. C. Martin, P. Kim,
H. L. Stormer and D. N. Basov, Nat. Phys. {\bf 4}, 532 (2008).


\bibitem{kram} C. Kramberger, R. Hambach, C. Giorgetti, M. H. R\"{u}mmeli, M.
Knupfer, J. Fink, B. B\"{u}chner,
Lucia Reining, E. Einarsson, S.
Maruyama, F. Sottile, K. Hannewald, V. Olevano, A. G.
Marinopoulos, and T. Pichler, Phys. Rev. Lett. {\bf 100}, 196803 (2008).

\bibitem{liu} Y. Liu, R. F. Willis, K. V. Emtsev, and
T. Seyller,
Phys. Rev. B {\bf 78}, 201403(R) (2008).

\bibitem{bost} A. Bostwick, T. Ohta, T. Seyller,
K. Horn and E.
Rotenberg, Nat. Phys. {\bf 3}, 36 (2007); A. Bostwick, T. Ohta,
J.L. McChesney, T. Seyller,
K. Horn and E. Rotenberg, Eur. Phys. J. Special Topics {\bf 148}, 5 (2007).


\bibitem{lu} J. Lu, K. P. Loh, H. Huang, W. Chen, and A. T. S. Wee ,
Phys. Rev. B {\bf 80}, 113410 (2009).

\bibitem{wangkl}K. L. Wang and D. M. Mittleman,
Nature (London) {\bf 432}, 376 (2004).

\bibitem{maie} S. A. Maier, S. R. Andrews, L. Mart\'{i}n-Moreno,
and F. J. Garc\'{i}a-Vidal,
Phys. Rev. Lett. {\bf 97}, 176805 (2006).

\bibitem{rana} F. Rana, IEEE Trans. Nanotech. {\bf 7}, 91 (2008).
\end{references}
\end{document}